\documentclass[referee]{raa}            

\usepackage{graphicx,times}             
\usepackage{amssymb}
\usepackage{mathrsfs}
\usepackage{amsmath}
\begin{document}

   \title{Possibility of Measuring spin precession of the nearest supermassive black hole by S stars
}

   \volnopage{Vol.0 (200x) No.0, 000--000}      
   \setcounter{page}{1}          

   \author{Wen-Biao Han}

   \institute{Shanghai Astronomical Observatory, 80 Nandan Road, Shanghai, 200030; {\it wbhan@shao.ac.cn}\\
}

   \date{Received~~2014 month day; accepted~~2014~~month day}

\abstract{ The supermassive black hole (SMBH) with a mass of 4 million $M_\odot$ inside the radio source Sgr A* in our Galactic center is the nearest SMBH. Once S stars with a shorter period are observed, the relativistic precessions especially the Lense-Thirring one can be measured by astronomical observations at 10 micro-arcsecond ($\mu$as) level in the future. An interesting but so far unaddressed problem is that the SMBH not only has spin but probably also has spin precession. We study the effect of such spin precession on the orbital precessions of orbiting stars. Our results show that the spin precession can produce a periodic oscillation on the precession of the star's orbital plane, but has no obvious effect on the periapse shift. For stars with an orbital period of $O(0.1)$ yr or less, such visible oscillations occur when the SMBH's spin-precession period ranges from about a few tens of years to hundreds of years. The period of oscillation is the same as the one of the spin precession. In principle, the precession of this oscillating orbital plane can be observed and then the spin and spin precession of the nearest SMBH can be determined.
}

   \authorrunning{W.-B. Han }            
   \titlerunning{Possibility of Measuring spin precession }  

   \maketitle

%
%
\section{Introduction}           
\label{sect:intro}

It is widely believed that there is a supermassive black hole (SMBH)
associated with the radio source Sgr A* in the center of our Galaxy
(\cite{kg01,bege03,shen05}). The mass of
supermassive black hole is estimated to be about 4 millon solar mass
(\cite{Ghez08,Gill09}). The distance from the SMBH to the Sun is about 8 kpc,
is 100 times closer than the SMBH in Andromeda, the nearest large
galaxy. For this reason, the Galactic black hole offers the best
laboratory for strong gravitational field physics and testing
general relativity (GR) (see reviews \cite{alex05,genz10} for more
details).

Using stars orbiting the Galactic SMBH, one can detect post-Newtonian effects
and test GR in the weak and strong field limits
near the SMBH. Several successful experiments have been done in our solar
system for detecting very weak GR effects
(\cite{kope03,Bert03,lucc10,Ever11}), but the SMBH will give us the
best chance to test GR in a strong gravitational field near an
SMBH. The star source 2 (S2), one of a group of stars labeled by S at
distances ranging from $10^0-10^2$ milliparsec (mpc) from the Galactic center,
has an orbital period about 15 years and an advancd of periapse of about
$0.2^\circ/{\textrm{yr}}$ based on general relativity. However, the
Lense-Thirring effect on S2 is too small to be detected with current technology. Recently,
Meyer et al. observed a star named S0-102 orbiting our Galactic SMBH with a period only 11.5 yr, which is the shortest one known (\cite{Meyer12}). As
suggested by Will (\cite{will08}), if we can find some stars around
the Galactic center at very small semimajors $a \lesssim 1
\textrm{mpc}$ and with high eccentricity, the orbital plane precessions
induced by the spin angular momentum $\boldsymbol{J}$ and quadrupole
moment $Q$ of the SMBH can be larger than $10 \mu
\textrm{as}/\textrm{yr}$ observing from the Earth and can be detected by
some upcoming projects, for example, GRAVITY.

However, we cannot exclude this kind of possibility: the spin axis of the SMBH is precessing. For example, the Earth has a period of spin precession about 25800 yr, and the angle between the rotation axis and precession one is about $23.5^\circ$. The origin of this precession mainly comes from the coupling between the $J_2$ of the Earth and gravitational force of the Sun and Moon. The sources of precession for the black hole could be the collision with a compact star, the relativistic geodetic precession induced by a mass orbiting the black hole, and the coupling between the quadrupole moment of the hole and the Newtonian gravitational field of external bodies. Therefore, we should consider the influences of the spin precession of the SMBH on the orbital evolution of the S stars. If the spin precession can produce some obvious effects on the orbits of S stars, researchers would be able to observe these effects and determine the spin precession of the SMBH. Otherwise, if we do not consider the spin precession, the observation maybe appear some ``anomalous'' phenomenon, and will influence data analysis which can obscure the test of general relativity and the features of the SMBH. In this cases, the assumption of the spin precession and corresponding research should be valuable.

In the present paper, we try to investigate the effects of the SMBH's spin precession on the orbital precession of a star inside  a radius of 1 mpc area from the Galactic center. Without losing any scientific generality, the star is treated as a test mass, and the dynamical equation includes the first and second post-Newtonian (1PN, 2PN respectively) corrections, frame-dragging effect and quadrupole moment of the SMBH. Actually, we do not know the precession angle and rate of the SMBH's spin axis. We select a ``rational'' range of the angle and rate, and numerically evolve the orbit of the star to extract the effects of the spin precession. This paper is organized as follows. In Section
2 we briefly introduce the relativistic precessions of a star orbiting the SMBH at the Galactic center. Then in Section 3, we calculate and analyze the orbital evolution of a star around the SMBH when including spin precession. In the last section, conclusions and discussions are made.


\section{Relativistic effects in the Galactic center}
\label{sect:effects}

As mentioned in Section 1, an SMBH with $M=4\times
10^6M_\odot$ is located in our Galactic center. The Schwarzschild radii
of the SMBH is about $0.08\textrm{AU}$, which covers $10\mu \textrm{as}$ as seen
from the Earth. If a star with semimajor axis $a$ orbits around the
SMBH, the orbital period is
\begin{equation}
P=\frac{2\pi a^{3/2}}{\sqrt{GM}}\approx 1.48 \tilde{a}^{3/2}
\,\textrm{yr},\label{period}
\end{equation}
where $\tilde{a}$ is the semimajor axis in units of mpc. From
Eq.(\ref{period}), we can see if a star has a semimajor axis of $0.1\sim
1\textrm{mpc}$, the period will be about $0.1\sim 1 \textrm{yr}$.
The orbital periapse and plane precessions per orbit are given as
(\cite{will08}),
\begin{align}
\Delta \bar{\omega}&= A_S-2A_J \cos i-\frac{1}{2}A_Q(1-3\cos^2i),\label{precession1}\\
\Delta \Omega &= A_J-A_Q\cos i,\label{precession2}
\end{align}
where $\Delta\bar{\omega}=\Delta\omega+\cos i \Delta \Omega$ is the
precession of pericenter relative to the fixed reference direction,
 $i,\,\omega$ and $\Omega$ are the orbital
inclination, argument of periapse, and longitude of ascending node,
respectively. $A_S, \,A_J$ and $A_Q$ represent the relativistic
effects due to the black hole's mass, angular momentum and
quadrupole moment, respectively. Where, 
\begin{align}
A_S&=\frac{6\pi}{c^2}\frac{GM}{(1-e^2)a}\approx
8.351(1-e^2)^{-1}\tilde{a}^{-1} \,\textrm{arcmin/yr},\label{effects}\\
A_J&=\frac{4\pi\chi}{c^3}\left[\frac{GM}{(1-e^2)a}\right]^{3/2}\approx
0.0769(1-e^2)^{-3/2}\chi\tilde{a}^{-3/2} \,\textrm{arcmin/yr},\label{effectj}\\
A_Q&=\frac{3\pi\chi^2}{c^4}\left[\frac{GM}{(1-e^2)a}\right]^{2}\approx
7.979\times10^{-4}(1-e^2)^{-2}\chi^2\tilde{a}^{-2}\,\textrm{arcmin/yr}.\label{effectq}
\end{align}
We can easily find that the Schwarzschild precession ($A_S$) is much larger
than the other two.

\begin{figure}[!h]
\begin{center}
\includegraphics[width=3.0in]{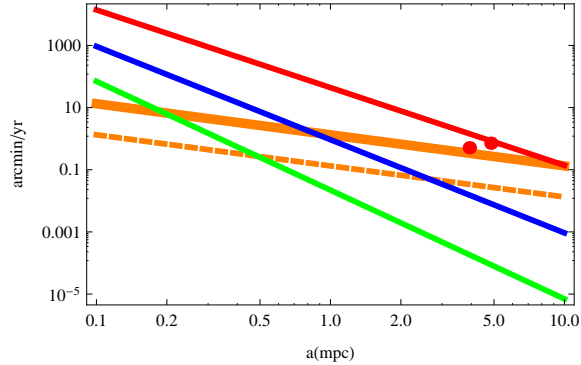}
\caption{The magnitude of precession for the Schwarzschild component ($A_S$, solid red line), angular momentum component ($A_J$, solid blue line) and quadrupole moment ($A_Q$, solid green line) compared with the precessions corresponding to observed astrometric displacements of $10\mu$as (solid orange line) and $1\mu$as (dashed orange line). The two red points label the S0-102 and S0-2 respectively.}\label{fig1}
\end{center}
\end{figure}

Figure \ref{fig1} plots the magnitudes of relativistic effects (\ref{effects})-(\ref{effectq}) (assuming $e=0.9$ and $\chi=1$). The solid and dashed orange bands denote the values of precession seen from the SMBH corresponding to astrometric precession rates of 10 and 1 $\mu$as per year as seen from the Earth respectively. We can see that for detecting the Schwarzschild part, the semi-major is required less than 10 mpc for a 10 $\mu$as observation accuracy from the Earth. Up to now, there are two S-stars have been found that satisfy this requirement: one is S0-2, and the other is S0-102, which has the shortest orbital period. The eccentricity of S0-102 is smaller than that of S0-2, so perhaps the relativistic effects of the latter one can be observed more easily. However, for both two stars, their orbital periods are larger than 10 years, which means that one needs tens of years of observations to measure the orbital precessions. For the frame-dragging effect, it only can be observed in stars that have an orbital period of $\lesssim 1$. S-stars with such a short period are also more convenient for assessing the Schwarzschild effect. An S-star with orbital radius $O(0.1 \text{mpc})$ is far away from the region where gravitational radiation is significant and also outside the tidal radius of the black hole (\cite{alex05}), therefore we do not need to consider these effects in the calculation.

\section{Effect of spin precession of the SMBH on S-stars' orbit}

For the completeness, the dynamical equation we used to calculate the motion of stars includes 1PN, frame-dragging effect, 2PN and quadrupole moment terms beyond Newtonian gravitation.
Considering the mass of SMBH is much larger than the orbiting stars,
the acceleration of the star can be written as,
\begin{align}
\boldsymbol{\ddot{x}}=\boldsymbol{a}_\text{N}+\boldsymbol{a}_\text{1PN}+\boldsymbol{a}_\text{2PN}+\boldsymbol{a}_{\text{F-D}}+\boldsymbol{a}_{\text{Q}}. \label{EoM}
\end{align}
We can see that in addition to the Newtonian part, 1PN and 2PN of the Schwarzschild part, the equation also includes the parts of spin: frame-dragging effect and quadrupole moment. From (\cite{Kidder95,will08}), we get the detailed expressions
\begin{align}
\boldsymbol{a}_{\text{1PN}}&=\frac{M\boldsymbol{x}}{r^3}(4\frac{M}{r}-v^2)+4\frac{M\dot{r}}{r^2}\boldsymbol{v},\\
\boldsymbol{a}_{\text{2PN}}&=-\frac{M\boldsymbol{x}}{r^3}\big(9\frac{M^2}{r^2}-2\frac{M}{r}\dot{r}^2\big)-2\frac{M^2}{r^3}\dot{r}\boldsymbol{v},\\
\boldsymbol{a}_{\text{F-D}}&=-\frac{2J}{r^3}[2\boldsymbol{v}\times\hat{J}-3\dot{r}\boldsymbol{n}\times\hat{J}-2\boldsymbol{n}(\boldsymbol{h}\cdot\hat{J})/r],\\
\boldsymbol{a}_{\text{Q}}&=\frac{3}{2}\frac{Q_2}{r^4}[5\boldsymbol{n}(\boldsymbol{n}\cdot\hat{J})^2-2(\boldsymbol{n}\cdot\hat{J})\hat{J}-\boldsymbol{n}],
\end{align}
where $\boldsymbol{x}$ is position vector of the star,  $\boldsymbol{v}$ velocity vector, $r$ the distance to the central black hole, $\boldsymbol{J}$ angular momentum of the black hole and $Q_2=-J^2/M$ the quadrupole moment based on the no-hair theorem.

Now for the purpose of this work, we need to add a term describing precession to the SMBH's spin axis. The precession equation is
\begin{align}
\boldsymbol{\dot{J}}=\boldsymbol{\Omega}\times\boldsymbol{J},\label{spinprecession}
\end{align}
where $\boldsymbol{\Omega}$ is the precession vector. As we know, the precession Eq.(\ref{spinprecession}) makes the spin axis rotates the precession axis but without changing the value of spin, that is, $|\boldsymbol{J}|=\text{const}$. The spin precession Eq.(\ref{spinprecession}) together with the equation of motion (\ref{EoM}) are used to calculate the S-stars' trajectories in this paper.


Now we estimate the possible value of the spin-precession frequency. Firstly, for the relativistic geodetic precession induced by a mass orbiting the SMBH, we have
\begin{equation}
\Omega_{\rm geodetic} \sim  (Gm_*/r^2c^2)(GM/r)\sim 10^{-9} {\rm yr}^{-1} (m_*/m_\odot)({r/\rm mpc})^{-5/2}, \label{precession1}
\end{equation}
where $m_*$ and $r$ is the mass and distance of the orbiting body. This looks extremely small. For example, a one solar mass star orbiting the SMBH from $1$ mpc will induce a precession with period of $10^9$ yr, which is almost the age of the Universe. The precession caused by the coupling between the quadrupole moment of the SMBH and the Newtonian gravitational field of external bodies is
\begin{equation}
\Omega_{\rm Quad} \sim  (3/2) (Q_2/J) (Gm/r^3) \sim  10^{-11} {\rm yr}^{-1} \, (\chi/1) (m_*/m_\odot) (r/{\rm mpc})^{-3}, \label{precession2}
\end{equation}
which is even less than the Eq.(\ref{precession1}) by several orders. Here $Q_2$ and $\chi$ are the quadrupole moment and Kerr parameter of the SMBH respectively.

Obviously, a $10^9$ yr spin precession is meaningless for research and observation. However, we cannot exclude that there is a star orbiting the SMBH in a distance of $\sim 0.1$ mpc radius, so the precession period could be $10^{6}$ years. And if inside the $0.01$ mpc area (still far from the tidal radius of the SMBH $\sim 3 \times 10^{-3}$), we can see the period reduces to ten thousand years. Furthermore, a compact object (neutron star or black hole with stellar mass) orbiting the SMBH on it's innermost stable circular orbit (this systems is called and extreme-mass-ratio inspiral, a very important source of gravitational wave for the eLISA project), the radius is  $\lesssim 10^{-3}$ mpc, which will make the central SMBH precess with a period of $\sim 10$ yr. Therefore, in this paper, we can assume a very large range values for the spin-precession frequency (or period) of the SMBH. We think that all these values could be realistic for this SMBH.

\begin{figure}[h!]
\begin{center}
\includegraphics[width=3.0in]{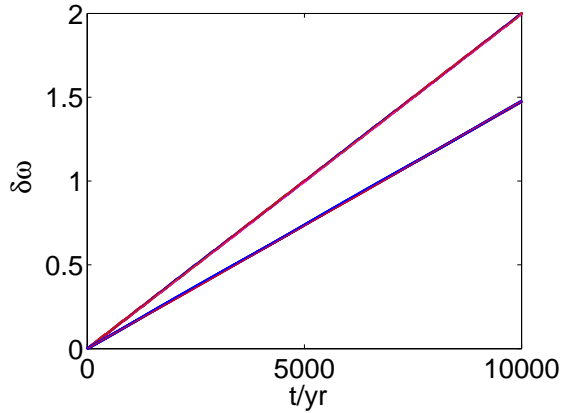}\caption{The periapse shifts of S0-102 (below line) and S0-2 (above one) for the spin-precessing SMBH. The precession angle is $60^\circ$, and precession frequencies are $0$, $10^{-10},~10^{-9},~\cdots,~10^{-1}$, and $0.5$ for each star.}\label{figs0s2w}
\end{center}
\end{figure}

Let's first see if the spin precession of the SMBH can affect the periapse shifts of known S stars S0-2 and S0-102. We set the precession angle to be $\pi/3$ and change the precession frequency $\varpi$ (unit $1/M$) to have values $10^{-10},~10^{-9},~10^{-8},~10^{-7},~10^{-6},~\cdots, ~10^{-1}$ and $0.5$ (corresponding precession period of: $\approx 4\times 10^4 \text{yr},~4\times 10^3 \text{yr},~4\times 10^2 \text{yr},~40 \text{yr},~4\text{yr},~\cdots,~21 \text{min}$ and 4.2 min.). The results are unfortunately no. In the Figure.\ref{figs0s2w}, and we only can see two lines represent the periapse shifts of S0-102 and S0-2. At this distance the spin precession of the SMBH has no obvious influence on the periapse shifts and then the lines denoting different spin precession overlap each other. This does not mean that the spin-precession has no couple with the periapse shift, but is quite small comparing to the magnitute of periapse shift. In fact the pericenter advance is dominated by the Schwarzschild part of the geometry, and the effect due to frame dragging is already a tiny correction, and any effect due to precession a tiny correction to that term. It is the reason that the curves in Figure. \ref{figs0s2w} don't show any effect of the spin precession.

\begin{figure}[h!]
\begin{center}
\includegraphics[width=2.8in]{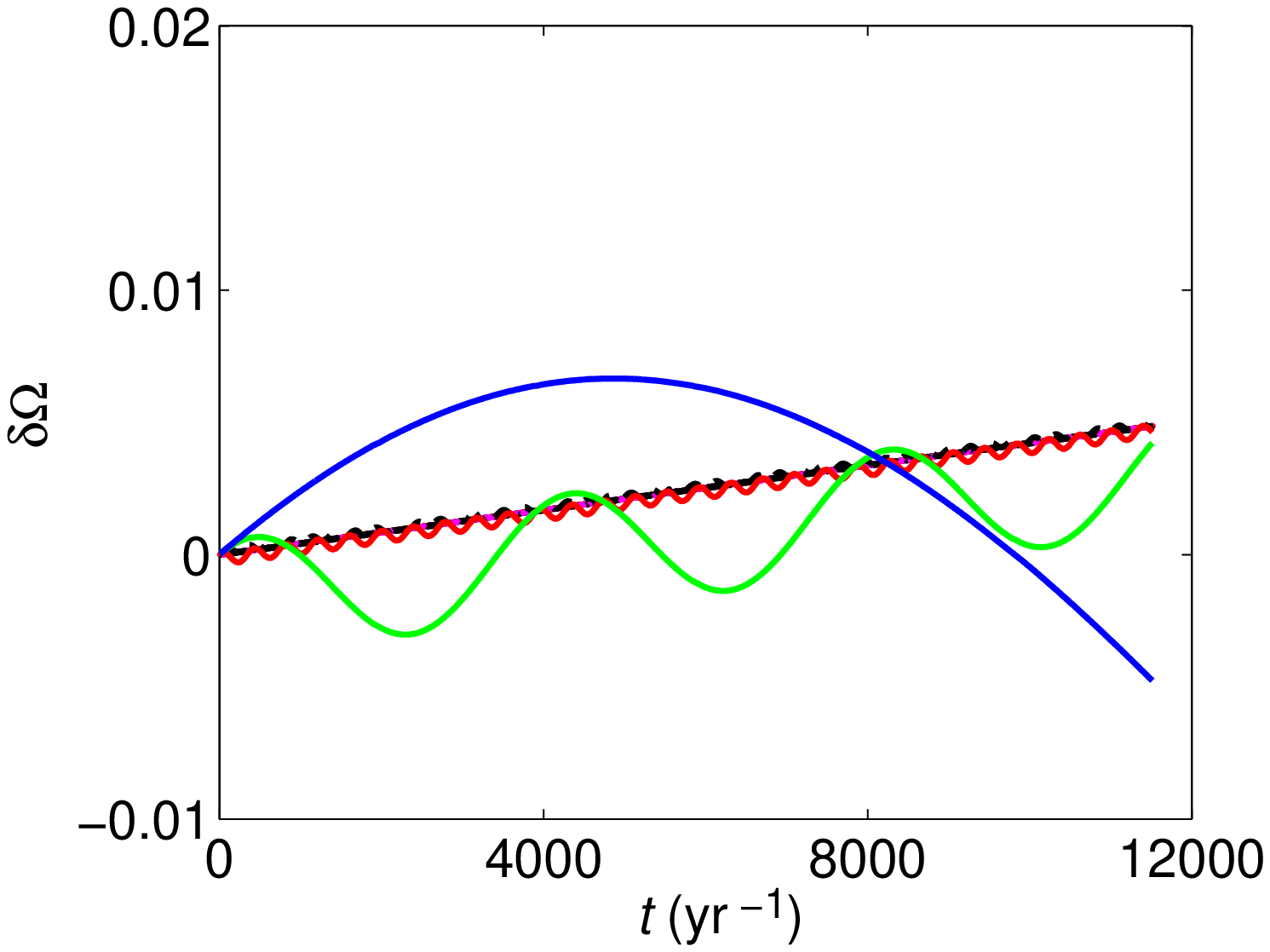}
\includegraphics[width=2.8in]{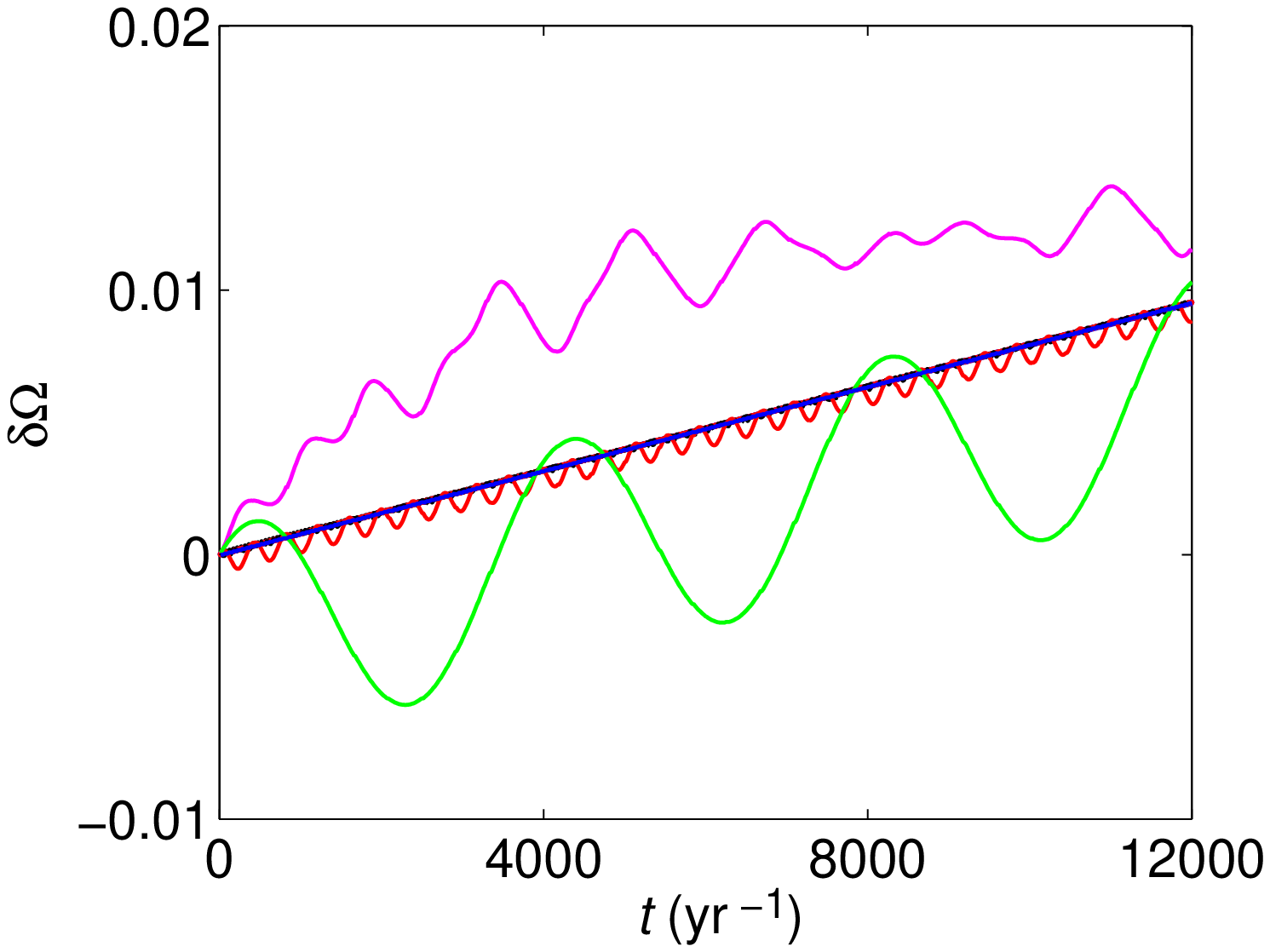}
\caption{The orbital precessions of S0-102 (left panel) and S0-2 (right panel) for the spin-precessing SMBH. The spin-precession parameters are as the same as Fig.\ref{figs0s2w}. The blue line labels $\varpi=10^{-10}$, the green line $\varpi=10^{-9}$ and the red line $\varpi=10^{-8}$. Attending the purple line in the right panel shows an anomalous effect for $\varpi=10^{-5}$. The other lines overlap and can not be recognized.}\label{figs0s2o}
\end{center}
\end{figure}

However, the effect of spin precession on the precession of orbital planes of S0-102 and S0-2 is obvious for several certain spin-precession periods. The SMBH's spin precession can make the orbital planes precess with a periodic oscillation behavior, and the oscillation period matches with the spin precession one (see Figure \ref{figs0s2o} for details). However, from Figure \ref{fig1}, we know that such frame-dragging effects of the both stars can not be observed even with a precision of $1\mu$as. Also the time scales are too long to observe these oscillations. Even for the periapse shift, because of the more than 10 years orbital period, it is also hard to observe.

\begin{figure}[h!]
\begin{center}
\includegraphics[width=2.8in]{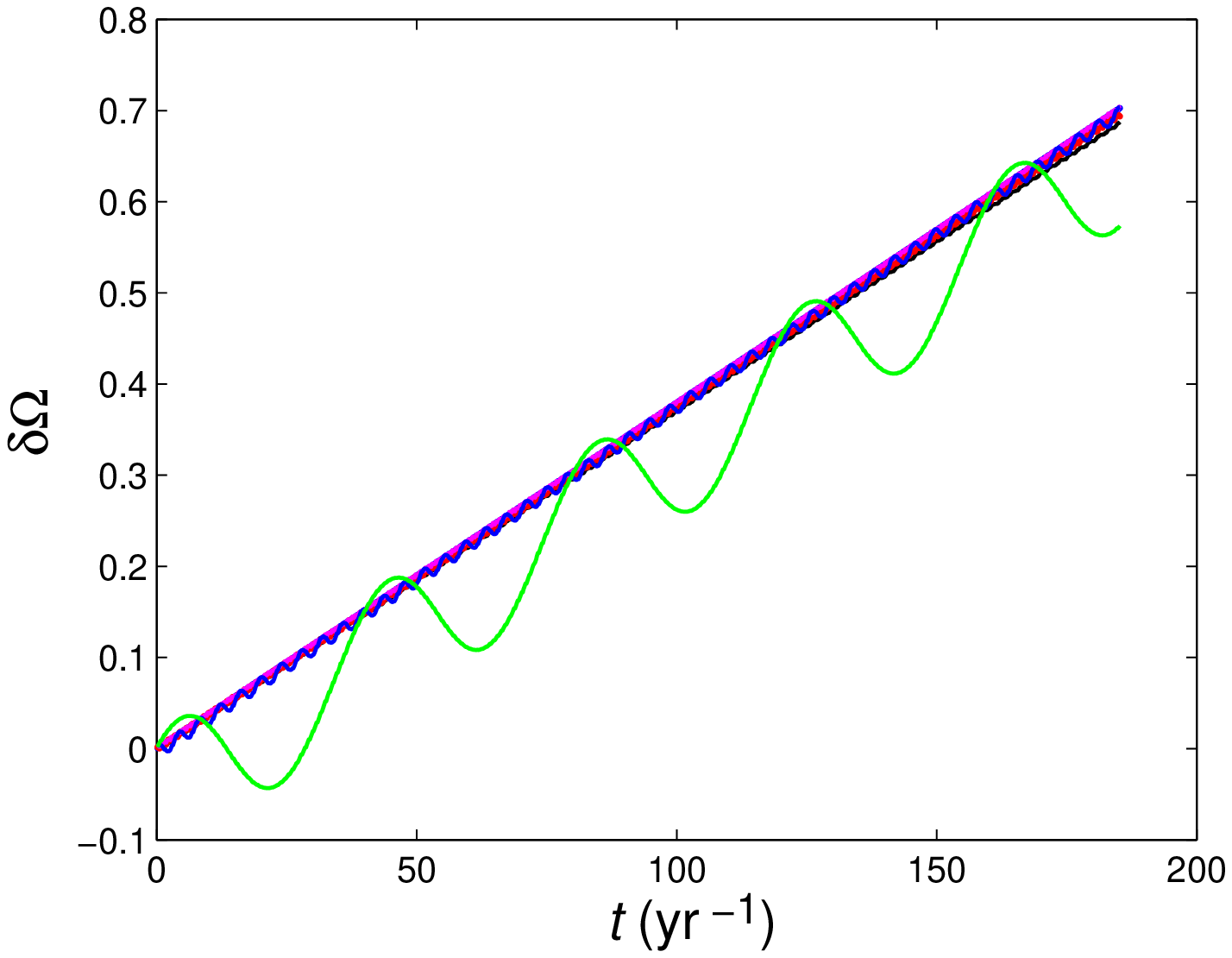}
\includegraphics[width=2.8in]{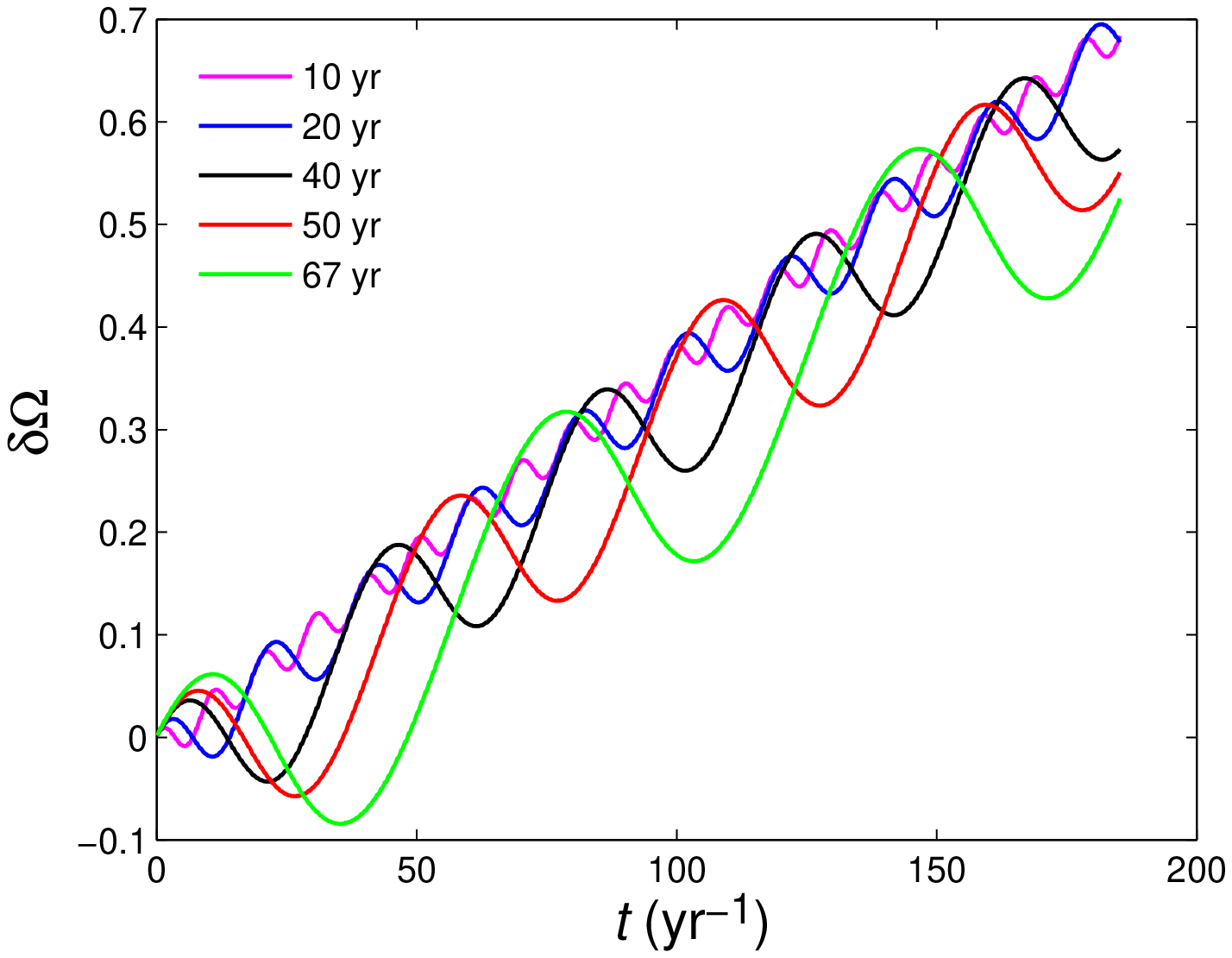}
\caption{The orbital precessions of a star with $a=0.3\text{mpc}, ~e=0.8$. Left panel: the spin-precession frequency from $10^{-7},~10^{-6},~\cdots,~10^{-1}$ and 0.5. The precession with frequency $10^{-7}$ (green line) has an obvious response on the orbital plane shift, and $10^{-6}$ also can be seen carefully (blue line). Right panel: several obvious oscillating phenomena.}\label{fig030}
\end{center}
\end{figure}

\begin{figure}[h!]
\begin{center}
\includegraphics[width=2.8in]{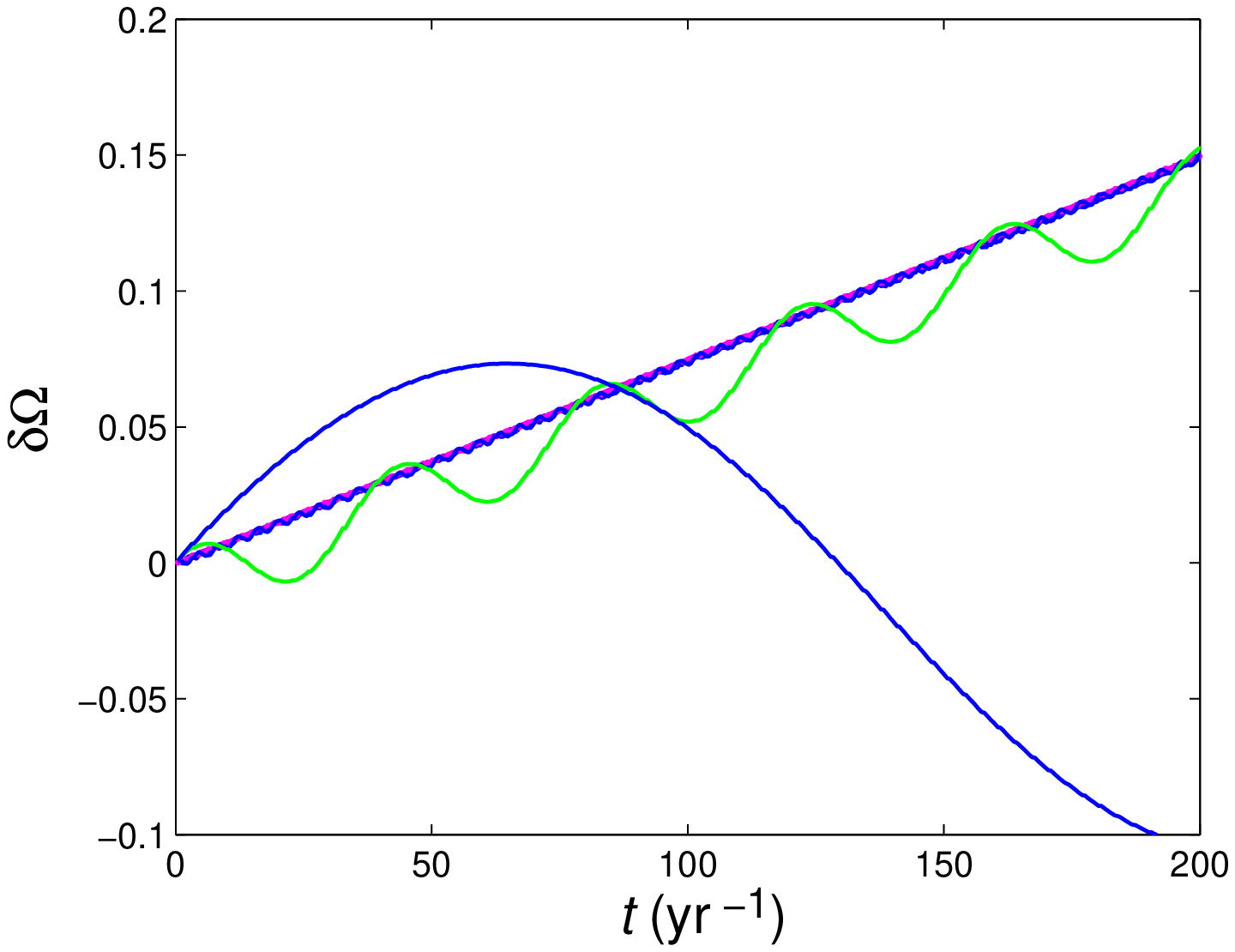}
\includegraphics[width=2.8in]{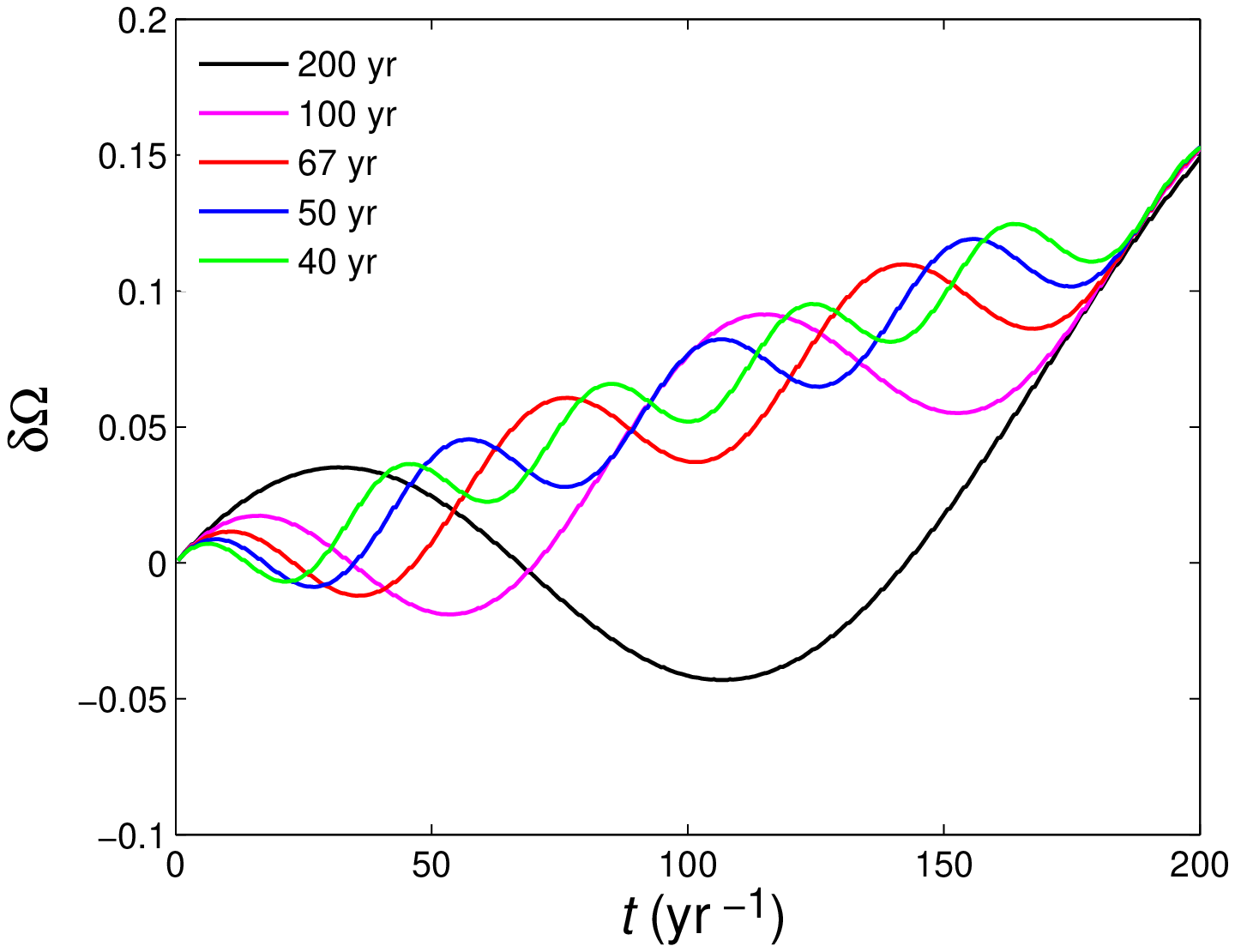}
\caption{The orbital precessions of a star with $a=0.5\text{mpc}, ~e=0.8$. Left panel: the spin-precession frequency from $10^{-8},~10^{-7},~\cdots,~10^{-1}$ and 0.5. The precession with frequency $10^{-8}$ (blue line) and $10^{-7}$ (green line) have obvious responses on the orbital plane shift. Right panel: several obvious oscillating phenomena.}\label{fig050}
\end{center}
\end{figure}

Let's now focus on the S stars with shorter orbital periods ($\sim O(0.1)$ yr), because as we have seen the relativistic effects of such stars can be measured at a level of precision of 10 $\mu$as. The calculation finds that the spin precession still has almost not obvious effect on the periapse shift. This time, the shift of orbital plane can be observed, and for certain spin precession, obvious oscillation behavior can be find with the same period of the spin precession. We show our results of two stars with $a=0.3$ mpc, $e=0.8$ and $a=0.5$ mpc $e=0.8$ respectively and the spin-precession angle is $30^\circ$, the response frequency of spin precession on the orbital plane evolution is around $10^{-6}$ to $10^{-8}$, which means, the oscillating period of the shift of orbital plane is around 10 yr to 400 yr. Observation can find the oscillation behavior if the SMBH has a spin precession with few ten years period. See Figures \ref{fig030} and \ref{fig050} for details.

The response of spin-precession period on evolution of the orbital plane reduces as the orbital period reduces. When the orbital period is ten years (like the S0-102 and S0-2), the spin precession with period thousands year can have an obvious effect on the shift of orbital plane; if the orbital period is $\sim O(1)$ yr, the corresponding spin-precession period is about a hundred years; And when the orbital period is $\sim O(0.1)$ yr, the corresponding spin-precession period is a few tens of years. We also calculate cases of anti-direction spin precession for these two stars, and find that the results are almost the same but the phases of oscillations is opposite with the ones demonstrated in Figures \ref{fig030} and \ref{fig050}.

\section{Conclusions and discussions}
\label{sect:discussion}
Our calculations demonstrate that a precession of the SMBH's spin axis can lead to an observable oscillation on the orbital plane precession of an S star. This responding oscillation has the same period as the spin precession. As is already known, for an astrometric precision of $10~\mu$as, observation of an S star with $O(0.1)$ yr orbital period can reveal the frame-dragging effect of the central black hole, and then the SMBH's spin. Furthermore, in this paper, we show that if the SMBH has a spin precession with a period of few ten years, the observation can also confirm such a spin precession. In principle, the spin of the SMBH interacts and couples with the orbit of the star, and change in the spin may trigger a variation in the orbit, then the advance of periastron. However for the distance scale we consider here, the periapse shift is mainly caused by the mass part of the SMBH, and at least two order larger than the frame-dragging effect. As a result, the process of spin-precession has no obvious effect on observation of advance of the periapse.

Such a precession period is possible for the SMBH in our Galaxy, as analyzed in Section 3. However, the period of the SMBH's precession is also possible longer or shorter. However the shorter or longer precession may not produce observable phenomena related to the star's orbital evolution. For the former one, though it can also cause the oscillating shift of orbital plane with the same period as itself, the amplitude of oscillation is too small to be observed. For the longer one, the time scale is too long to observe the oscillation behavior in a decade.

For the spin-precession value in this paper, we assume a compact star out the ISCO of the SMBH to make the spin-axis precession. One may ask that if this compact star with solar mass will perturb the target star and pollute the observation of orbital precession. We have calculated the perturbing effect of a close star on the target one in a previous paper (\cite{han12}), and we found that when the distance of two stars are only 0.01 mpc, a solar mass perturbation effect are much less (only 1\%) than the frame-dragging one. In this paper, the distance of two stars is much larger (the target star's semimajor is 0.3 -0.5 mpc, but the perturbed star's one is only about 0.001 mpc). In this sense, we can absolutely omit the perturbation of such a compact star.

\begin{acknowledgements}
This work was funded by the National Natural Science Foundation of China (NSFC) under No.11273045. We thank the very useful comments from the anonymous referees.
\end{acknowledgements}

\label{lastpage}

\end{document}